\documentstyle[12pt,thmsa,sw20lart]{article}

\input{tcilatex}
\begin{document}

\author{{\Large S. K. Nikogosyan , A. A. Sahakyan , H. N. Yeritsyan} \\
{\it Yerevan Physics Institute, Radiation Physics Laboratoty, }\\
{\it 2 Alikhanian Brothers St., 375036, Yerevan, Armenia\bigskip \medskip }
\and {\Large V. A. Grigoryan} \\
{\it Yerevan State University, Department of Physics, }\\
{\it 1 Alec Manoogian St., 375049, Yerevan, Armenia}}
\title{{\Large The influence of the bulk density on the intergranular properties of
YBa}$_{2}${\Large Cu}$_{3-X}${\Large Fe}$_{X}${\Large O}$_{Y}${\Large (0}$%
\leq ${\Large x}$\leq ${\Large 0.01)\ ceramics }}
\date{Preprint YerPhi \# 1519(19)-98}
\maketitle

\begin{abstract}
The influence of the bulk density of YBa$_{2}$Cu$_{3-X}$Fe$_{X}$O$_{Y}$ (0$%
\leq $X$\leq $0.01) ceramics on the intergranular superconducting (SC)
properties was studied using the temperature dependence of AC magnetic
susceptibility measurements. It was found that the simultaneous variation of
the sample's density and the iron impurity concentration does not influence
effectively the onset temperature of the superconducting state $T_{c}^{on}$.
While only increasing of the sample's density shifts the intergranular
hysteresis losses peak temperature $T_{m}^{J}$ to the lower values which
connects with the decreasing of the Josephson magnetic vortices pinning
role. It was established that the shielding capability and $T_{m}^{J}$
display a plateau with $X$ in the $0.003\leq X\leq 0.007$ region which is
due to the monotonous decrease of the sample's density. It was shown that
the shielding capability at the $T=78K$ for the sample with $3.8g/cm^{3}$ is
two times higher than that for the sample with the density of $5.0g/cm^{3}$.
The possible interpretations of the observed results are discussed.
\end{abstract}

\section{Introduction}

About the crucial role of bulk density $(\rho )$ in the formation of SC
properties in the oxide alloys was discussed in refs. [1-9]. The bulk
density $\rho $ was defined by sintering pressure and by choosing the
pressure regime and also by temperature and annealing time of the sintering
[1-4].

At the same time, the decrease of the granule sizes and the increase of the $%
\rho $ and further saturation was observed when the sintering pressure
reaches the limiting value \cite{1,2}. It was shown that the pressing rate
is a more critical parameter than the value of sintering pressure in the
formation of the SC and structural properties of the obtained samples [5].
In ref. \cite{6} was observed a decrease of the oxidation rate with the
increase of density of $YBa_{2}Cu_{3-X}Fe_{X}O_{Y}$ ceramics which makes it
difficult to grow a sample with a higher $X$. An increase of critical
current $(J_{c})$ and magnetic susceptibility$(\chi ^{\prime })$ in the
superconducting state and also a decrease of resistivity of the sample in
normal state was observed with increasing of $\rho $ in [1,2,7,8]. While in
ref. \cite{3} the samples with higher $\rho $ have lower $J_{c}$ . On the
other hand , it was established that the onset temperature $T_{c}^{0}$ of
the resistivity appearance increases , and the onset temperature for
diamagnetism $T_{d}^{on}$ decreases with the increasing of $\rho $ in Y- and
Bi-based samples \cite{7,8}.

Note, also that the critical transition temperature $T_{c}$ in Bi-based
samples remains unchanged [2] , although their density $\rho $ varies in a
significantly large range. At the same time the critical temperatures,
defined by onset, midpoint and offset of SC transition, depending on
sintering pressure (or on density $\rho $), display nonmonotonic behavior 
\cite{4,5}. In $Bi_{3.2}Pb_{0.8}Sr_{4}Ca_{5}Cu_{7}O_{Y}$ samples the
hysteresis losses peak $(\chi ^{\prime \prime })$ temperature $T_{m}^{J}$
and onset temperature of diamagnetism $(\chi ^{\prime })$ $T_{d}^{on}$
shifts to the lower values with increasing of their $\rho $ [8].

It was shown \cite{9}, that the temperature of complete penetration of high
frequency $($ $\sim 4$ $MHz)$ electromagnetic field into SC $YBaCuO$ plates $%
T_{m}^{J}$ shifts to the lower value region with increasing of $\rho $, in
contrast with [8]. Thus the analysis of literature data shows that the role
of $\rho $ in formation of physical properties of HTSCs is important and it
is not clarified unambiguously at present.

In this paper using AC susceptibility measurements we present the results of
study of intergranular SC properties of the $YBa_2Cu_{3-X}Fe_XO_Y$ $(0\leq
X\leq 0.01)$ ceramics with fixed and variable compositions depending on $%
\rho $.

\section{Experimental}

An AC susceptibility $\chi ^{\prime }-i\chi ^{\prime \prime }$ measurement
technique was applied in order to determine the critical parameters of high $%
T_{c}$ superconductors. The experimental equipment is described in detaile
in our previous papers \cite{10,11}. Note that the sample's temperature
changed from 78 to 100 K and was measured within the accuracy of $\pm $ 0.1
K.

The samples of about 3 mm in diameter and about 9 mm in height were prepared
by the well known ceramic technology. The concentration $X$ of $Fe^{57}$
isotops which substituted copper atoms were: 0; 0.002; 0.003; 0.005; 0.007
and 0.01 which are subsequently referred as \# 0, 1, 2, 3, 4 and 5. For
these samples the $T_{c}^{on}$ was changed between 91 and 92 K, and their
bulk density had the following values: 5.29; 5.22; 4.64; 4.71; 3.92 and 5.00 
$g/cm^{3}$, respectively. We have also studied another sample \#6,
containing atom of $Fe^{56}$ isotope with $X=0.01$ as a substitutional
element for copper and having the following parameters:$\rho =3.8g/cm^{3}$ ; 
$T_{c}^{on}=91.5$ $K$. The sample's density was defined by their weights and
sizes. Note that we does not measure the peak positions of $\chi ^{\prime
\prime }$ for sample \#5 because they were below 78 K [11].

Note, that in our investigation we also use the shielding capability of the
samples. As such a parameter we have chosen the value of $\chi ^{\prime }$
at 78 K in arbitrary units.

\section{Results and discussions}

Fig. 1 shows the temperature dependences of imaginary $\chi ^{\prime \prime
} $ (a) and real $\chi ^{\prime \prime }$ (b) parts of magnetic
susceptibility for the samples with variable composition and different
density (curves from 0 to 5). The analogous results are presented in the
same fig. (curves 5 and 6) for samples \#5 and \#6 with the same composition
X = 0.01 and different densities 5.00 and 3.87$g/cm^{3}$, respectively.

As is seen in fig.1 with the increasing of X a shift of both the $T_{m}^{J}$
peak position on the $\chi ^{\prime \prime }(T$ $)$ curve and the
corresponding low temperature inflection point on the $\chi ^{\prime \prime
}(T$ $)$ curve to the low temperatures was observed which agrees with our
previous results [11]. Such a behavior display also the samples having $%
X=0.01$ with increasing of $\rho $. At the same time the broadening of the
SC transition was observed on both curves $\chi ^{\prime }(T$ $)$ and $\chi
^{\prime \prime }(T$ $)$. The strong dependence of SC properties versus
density is obvious from fig. 1 where the $T_{m}^{J}$ for sample \#5 was
shifted to below the temperature of liquid nitrogen while for sample \#6 $%
T_{m}^{J}$ $=$ $88.8$ K . This behavior agrees with our previous results of $%
T_{m}^{J}$ versus X and it can be explained as follows. The $T_{m}^{J}$
decreasing with increasing of X is due to the formation of weak links in the
granuls [11,12]. And the observed decrease of $T_{m}^{J}$ with increase of
sample's density , as assumed in [7,8], may be caused by the reduction of
the role of Josephson vortices pinning which is in its turn due to the
reduction of the thickness of intergranular layers. It can be mentioned that
the high density samples sintered by conventional method, have low oxygen
content [6] which enhances the role of inhomogeneities induced in samples by
iron doping \cite{11,12,13,14}. Hence, these inhomogeneities lead to the
additional broadening of the SC transition in denser samples (fig.1 , curves
3 and 4 ; 5 and 6).

One can also indicate a peculiarity of $\chi ^{\prime }(T$ $)$ curves: the
shielding capability (fig. 1. B; 5 ,6 curves) decreases twice with the
increase of $\rho $ for the fixed value $X=0.01$ which contradicts the
results in [7,8]. This indicates additionally that the samples with higher
density can have an oxygen deficiency [6] which cause the reduction of $\chi
^{\prime }$ \cite{15,16}.

Note, that the $T_{c}^{on}$ versus $\rho $ shows a fluctuation character and
ranges in 91 - 92 K interval. Such behavior was found in undoped Y- based
samples [4,5] with the difference that the $T_{c}^{on}$ changes monotonously
from 89 to 95 K with increasing of $\rho $ from 3.4 to 5.4 $g/cm^{3}$. The
variation of $T_{c}^{on}$ with $\rho $ according to [4,5] can be explained
as follows. The samples with higher density are sintered under the high
pressure and have a high concentration of structural defects, created by
plastic deformation after the pressure. These defects are probably
concentrated near the grain boundaries and reduce strongly the $T_{m}^{J}$,
which is observed in our experiments for the \#5 sample having higher
density in comparison with sample \#6. If the accumulated structural defect
concentration exceeds some critical value, the sample undergoes from
orthorhombic to the tetragonal type of symetry [4,5]. Hence in order to
obtain samples with higher density and better SC properties it required that
they undergo special thermal treatment [5].

From the magnetic field dependence of SC transition in sample one can
speculate about pinning of magnetic flux lines, which leads to gaining
information about the HTSC phenomenon [11,12]. The run of $\chi ^{\prime
\prime }(T$ $)$ curves for some applied AC magnetic field amplitudes in
sample \#3 are presented in fig. 2. And curves of $T_{m}^{J}(h_{0})$ for all
samples are also presented in fig. 3. The common behavior for all samples is
the shift of $T_{m}^{J}$ to the lower values with increasing of magnetic
field amplitude $h_{0}$. It is typical that all curves $T_{m}^{J}(h_{0})$
reveal strong and weak region of decreasing of $T_{m}^{J}$ which evidences
about the existence in samples of ''strong'' and ''weak'' pinning centres
for Josephson magnetic vortices [11]. From figures 2, 3 one finds that the
magnetic field $h_{0}$ region for strong decreasing of $T_{m}^{J}$ is
narrowing with increasing of $\rho $ which agrees with the results obtained
from the measurements of high frequency electromagnetic field absorption in
Y-based samples [9]. However it should be emphasized that the latter, well
agrees with the results of refs. [7,8] , but condraticts ref. [9]. The
reason for the discrepancy is not obvious , however , it may be attributed
to the different techniques for detecting of critical parameters in referred
works. For instance, in Y - based samples the onset temperature of
diamagnetism $T_{d}^{0}$ decreases [7,8], while the onset temperature of
appearance of the electroresistivity, increases with increasing of $\rho $
[7].

As it is known [9,11], the slope of $T_{m}^{J}(h_{0})$ curve characterizes
the pinning force for Josefson vortices and the higher the slope is the
lower the pinning force becomes. One can find from fig.3 that for sample \#3
with lower iron content and higher $\rho $ in respect to sample \#6, the
curve $T_{m}^{J}(h_{0})$ at $h_{0}\prec $ $0.1$ $Oe$ has a higher slope.
Hence, the increase of sample density leads to the decrease of pinning force
and/or to decrease of the pinning role of Josefson vortices [7,8].

In \#2, 3, 4 samples the iron content $X$ increases monotonously while their
densities almost monotonously decrease. Hence the dropping rate of $%
T_{m}^{J}(h_{0})$ in these samples consequently decreases. Fig. 4 represents
both the dependence of $T_{m}^{J}$ for several $h_{0}$ and the shielding
capability at T = 78 K versus iron concentration which clearly describes the
role of bulk density in variation of SC properties of samples. It is obvious
that these two parameters display the correlative behavior: they sharply
drop with initial increasing of X and formate a plateau in $0.003\leq X\leq
0.007$ region and further decrease when $X$ reaches to 0.01(sample \#5) .The
observed plateau can be explained as follows. In the first two samples the
density almost does not change, only the iron concentration $X$ increases
which leads to the decrease of $T_{m}^{J}$ [11,12]. Further, although the $X$
increases, the $\rho $ almost monotonously decreases which according to
[7,8] leads to the decrease of droping rate of $T_{m}^{J}$ versus $X$ and
hence causes the formation of a plateau in the above mentioned $X$ region.
Further, the simultaneous increase of $X$ and $\rho $ leads to the strong
shift of $T_{m}^{J}$ to the temperatures below 78 K [8], as seen from fig. 1
(curve 5). Therefore in fig. 4 the imaginary run of $T_{m}^{J}$ for $X=0.01$
is presented by dotted lines.From this figure one can see that the sample
\#6, even for relatively higher values of $h_{0}$, has significantly higher
values of $T_{m}^{J}$ (fig.4, symbols 4 and 5). This means that in the
samples with higher density the $T_{m}^{J}$ drops rapidly with increasing $X$%
.

\section{Conclusion}

The analysis of intergranular SC properties of $YBa_{2}Cu_{3-X}Fe_{X}O_{Y}$
ceramics with different $X$ and bulk density allows to draw the following
conclusions.

1. The onset transition temperature to the superconducting state $T_c^{on}$
shows a fluctuation character and varies from 91 to 92 K with increasing
bulk density and iron concentration.

2. The intergranular losses peak temperature $T_{m}^{J}$ decreases rapidly
with the increasing of iron content in the samples with a higher density.
Partialy, in the samples with $X=0.01$ and higher density the shift of $%
T_{m}^{J}$ to the region below 78 K was observed.

3. The simultaneous increase of iron concentration and bulk density in the
samples leads to the rapid decrease of $T_{m}^{J}$. However, a plateau is
observed on the $T_{m}^{J}(X)$ curves in the $0.003\leq X\leq 0.007$ region
due to the almost monotonous decrease of density in the same region.

4. The shielding capability at 78 K also shows a plateau in the same $X$
region. Besides, it was found that this parameter is about twice lower for
the samples with fixed $X=0.01$ and higher density, which contradicts the
results obtained in refs. [7,8]. Such behavior probably may be attributed to
the oxygen deficiency in the samples due to the higher bulk density.

\FRAME{ftbpFU}{4.5887in}{5.0055in}{0pt}{\Qcb{ Dependence of intergranular
losses peak temperature $T_{m}^{J}$ versus AC magnetic filed amplitude $%
h_{0} $ in samples. The curves from 0 to 6 correspond to \#1, \#2, \#3, \#4
and \#6 samples, respectively.\protect\bigskip }}{\Qlb{Fig.3}}{fig3.eps}{%
\special{language "Scientific Word";type "GRAPHIC";maintain-aspect-ratio
TRUE;display "USEDEF";valid_file "F";width 4.5887in;height 5.0055in;depth
0pt;original-width 439.625pt;original-height 479.8125pt;cropleft "0";croptop
"1";cropright "1";cropbottom "0";filename 'C:/My
Documents/Parts/xtutun/fig3.EPS';file-properties "XNPEU";}}

\FRAME{ftbpFU}{3.9461in}{5.2209in}{0pt}{\Qcb{ Dependence of intergranular
losses peak temperature $T_{m}^{J}$ for several values of $h_{0}$ (1, 2, 3,
4, 5 curves) and shielding capability at 78 K (6) versus iron concentration
X. 1, 2 and 3 - $T_{m}^{J}(X)$ curves for applied AC magnetic field
ampltudes $h_{0}$ (Oe): 0.025, 0.1 and 0.3, respectively; 4 and 5 - $%
T_{m}^{J}$ values in sample \#6 , respectively for $h_{0}=0.1Oe$ and $%
h_{0}=0.3Oe.$}}{\Qlb{Fig.4}}{fig4.eps}{\special{language "Scientific
Word";type "GRAPHIC";maintain-aspect-ratio TRUE;display "USEDEF";valid_file
"F";width 3.9461in;height 5.2209in;depth 0pt;original-width
333.25pt;original-height 441.625pt;cropleft "0";croptop "1";cropright
"1";cropbottom "0";filename 'C:/My
Documents/Parts/xtutun/fig4.EPS';file-properties "XNPEU";}}

\end{document}